# Language Support for Optional Functionality

Joy Mukherjee, Srinidhi Varadarajan
*660 McBryde Hall,*
*Dept. of Computer Science*
*Virginia Tech,*
*Blacksburg VA 24061*
*1- (540)-231-9431*
*{jmukherj, srinidhi}@vt.edu*

## Abstract

We recommend a programming construct – availability check – for programs that need to automatically adjust to presence or absence of segments of code. The idea is to check the existence of a valid definition before a function call is invoked. The syntax is that of a simple 'if' statement. The vision is to enable customization of application functionality through addition or removal of optional components, but without requiring complete re-building. Focus is on C-like compiled procedural languages and UNIX-based systems. Essentially, our approach attempts to combine the flexibility of dynamic libraries with the usability of utility (dependency) libraries. We outline the benefits over prevalent strategies mainly in terms of development complexity, crudely measured as lesser lines of code. We also allude to performance and flexibility facets. A Preliminary implementation and figures from early experimental evaluation are presented.

## 1 Introduction

Application software such as browsers, email clients, and media players continue to increase in size and complexity. This phenomenon has fueled the need for customized installations where many features are made *optional modules*. Given the unpredictability and variety of user-needs, installations may have to be frequently re-customized as well. Furthermore, embedded processors in handhelds and cellular telephony are on the rise. An important characteristic of these devices is that they have relatively small memory and storage footprints. Such constraints manifest as increased necessity for customization. To quote early examples, we modified DILLO (a web-browser for embedded environments) [2] to automatically adjust to optional support for images of various kinds [Figure 1]. The modified version involved 3 optional modules of combined size 35% of the overall application. XMMS [12], a media player for Linux, comprises 20 optional pluggable modules of total size 40% of overall code.

The issue of customization is particularly pertinent to the domain of compiled software. The complexity of tailored compilation makes it a non practical solution for regular users. Sources of proprietary software are not even available for local compilation. In this regard, mechanisms that facilitate automatic adjustment to removal (or addition) of code pieces from (to) compiled applications can be useful. For instance, an undesired functionality can be removed simply by deleting the associated code module. Similarly, a non-available feature may be reinstated by restoring the corresponding module. Again many companies provide pieces of software as modules to be plugged into a main application developed by others. Differing tenets of involved institutions and related security concerns can only be bridged through automatic post-building mechanisms.

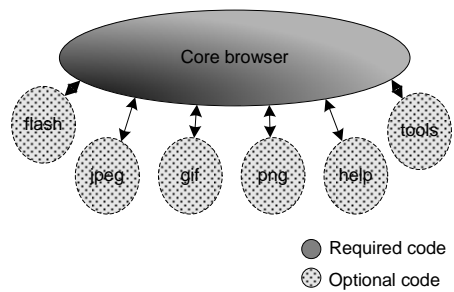

Figure 1: A customizable web-browser can comprise optional support for images, flash, help, and other tools.

In this paper, we recommend a programming construct – availability check – for programs that need to automatically adjust to presence or absence of segments of code. Focus is on C-like compiled procedural languages and UNIX-based systems (section 2). Essentially, our approach attempts to combine the flexibility to dynamic libraries with the usability of utility (dependency) libraries (sections 3, 4, 5). We detail on preliminary implementation (section 6), and discuss alternatives (section 7). Benefits over prevalent strategies are presented mainly in terms of development complexity, crudely measured as lesser lines of code (section 8). Lastly, we outline ongoing work (section 9).

## 2 Motivation

Many contemporary applications and utility software continue to use C-like compiled procedural languages. Both DILLO and XMMS are examples. Compilation scores over interpretation and JIT technologies [5] in terms of performance, but loses out on flexibility. While prevalent mechanisms for application-controlled dynamic loading and linking can aide automatic adjustments, they require special programming constructs. This increases programmer burden and program complexity. Our objective is to reduce such programming complexity. We focus on UNIX-based systems mainly because of our own familiarity with them and greater access to system internals. Nevertheless, we believe that the import of this work is valid beyond C and UNIX.

## 3 Shared Objects

UNIX-based systems provide for shared object files [9] as loadable modules external to a main program. Typically, shared objects are used in two distinct manners by C programs (C++ follows similar semantics).

### 3.1 Dynamic Libraries

Shared objects can be used as dynamic libraries [4][9] for on-demand runtime extension of program functionality. Applications can control loading and linking of dynamic libraries using services in `libdl` [3]. Optional

```
void *handle;
void (*foo)(void);
void container_function ()
{
    handle = dlopen (libfoo);
    if (handle)
    {
        foo = dlsym (handle, "foo");
        if (foo) // check for NULL
            (*foo) ();
    }
    dlclose (libfoo);
}
```

Figure 2: Loading a dynamic library (libfoo) and linking to a function (foo) in it entails explicit programming.

features, built into dynamic libraries, can be instated in a specified directory to be made available. Similarly, their removal is achieved through purging of libraries from a directory. Dynamic libraries are frequently used to implement plugins for C programs. Our base-versions of DILLO and XMMS use this technique.

Nonetheless, loading, unloading, and linking of dynamic libraries has to be explicitly programmed into an application [Figure 2]. This results in extra lines of code as well as knowledge of special constructs (`dlopen`, `dlsym`). Our contention is that alleviating such programming complexities will ease the increasing use of shared objects for optional features.

### 3.2 Utility Libraries

Shared objects are also frequently used as utility or dependency libraries [1][7] that implement basic and widely used services. Utility libraries prove effective in decoupling utility services from a particular program, thus making them available to one and all [Figure 3]. XMMS and DILLO have 20 common dependency libraries, totaling at least 1.5 times the size of either executable. Besides, using utility libraries is easy. Once bound to a program using compile-time linker flags (`-l` for `gcc` on UNIX), they are automatically loaded and linked in from specified locations. They do not entail use of special programming and their services can be accessed as conventional external functions or variables from inside core application code.

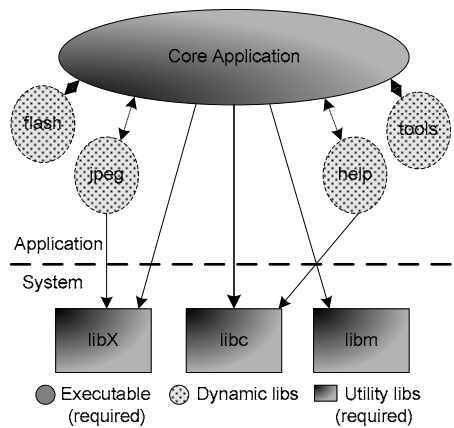

Figure 3: Dynamic libraries are used for optional application modules. Utility libraries offer common services.

Currently however, utility libraries are used for *necessary* codes and must be loaded at start-up. Their non-availability is fatal to an application. Our contribution lies in alleviating the critical necessity of utility libraries thus allowing their use for optional extensions. We attempt to combine the flexibility of dynamic libraries with the usability of utilities.

## 4 Function Availability Check

The programming construct central to this work involves a runtime availability check on a function before it is called. The call is made if and only if the availability check succeeds i.e. a definition for the function is found. Syntactically, function availability check (FAC) is a simple 'if statement' [Figure 4].

```
extern void foo ();
void container_function ()
{
    if (foo)
        foo ();
}
```

Figure 4: Syntax of FAC is like a simple 'if'.

The intuition is that an available or defined function has a valid non-NULL address associated with it while an undefined function does not. Note that, unlike variables (including those of type function pointer), an undefined function should not be misconstrued to have an address with a NULL value. There is simply *no notion of an associated address with respect to undefined functions*. Naturally, FAC is pertinent to external functions only, since internal functions are available by definition. Note that FAC is a type less construct. This is because, conceptually, the type of a function is unimportant unless it is available. Moreover, from an implementation perspective, it may be impossible to check the type of a function that is not available. Lastly, FAC must be performable on functions of all type and hence, is orthogonal to typing.

## 5 Optional Utility Libraries

Function availability checks can aid automatic adjustments to presence or absence of utility libraries. Consider the pseudo C code depicted in [Figure 5]. It is based on DILLO's handler for MIME types such as text, html, images etc. The portable network graphics (PNG) [8] handler is optional and may be dispensed with during compile-time (preprocessing) configuration.

```
…
extern Png_handler_fn (…);
…
void Mime_handler (…)
{
#ifdef ENABLE_PNG
    if (mime_type == "image/png")
        Png_handler_fn (…);
#endif
    …
    if (mime_type == "text")
        Text_handler_fn (…);
}
```

Figure 5: DILLO handler for MIME types. The handler for text is core, while that for PNG is optional.

Now consider a FAC-based version of the same code [Figure 6]. Here, availability of the PNG handler is checked at runtime and the function is called only on success. If the program loader is lenient towards absence of the concerned utility library at startup (see section 6.2), this version of the program can automatically adjust to optional PNG support. More importantly, FAC maintains the semantics of a simple 'if statement'.

```
…
extern Png_handler_fn (…);
…
void Mime_handler (…)
{
    …
    if (Png_handler_fn)
        if (mime_type == "image/png")
            Png_handler_fn (…);

    if (mime_type == "text")
        Text_handler_fn (…);
}
```

Figure 6: FAC-based handling of optional PNG MIME.

In typical 'if'-like fashion, FAC can also be used for complex structures such as compound statements and if-else cases. A compound FAC statement can be useful in isolating several subsequent statements that are contingent on the target function. Furthermore, it may be used to avoid multiple FACs on functions contained in the same library. 'if-else' cases of FAC can, possibly, be effective in availability-driven runtime selection between alternate ways to achieving some functionality.

## 6 Implementation

This section discusses how to address implementation challenges on an x86 platform running GNU/Linux. The implementation comprises (1) compiler modifications to allow FAC, and (2) linker/loader modifications to inculcate tolerance for missing utility libraries. Most flavors of GNU/Linux use the Executable and Linkable Format (ELF) for compiled binaries [11]. Consequently, the implementation relies on ELF concepts.

### 6.1 Compilation

Normally, contemporary procedural programming does not allow undefined references. (Weak references are discussed in section 6.3). Also, as mentioned earlier, a defined function reference must have a valid non-NULL address. Consequently, traditional program semantics consider function availability checks to be redundant and compilers remove them to optimize the code. However, this phenomenon implies that FAC-like statements are recognized as distinct from other regular 'if' statements. Thus incorporating FAC support into a compiler such as gcc does not necessitate modifications to the parser, grammar, or state machine. It may be achieved through 3 modified compiler steps – (1) on encountering a seemingly redundant if-statement, check if the involved *symbol* is a function. (2) If the symbol is not a function, take traditional action. (3) If the symbol is a function, generate FAC code [Figure 7].

Recall that FAC needs to be performed on external functions made available through utility libraries. On our target platform, external function calls from a program to utility libraries are redirected through a Procedure Linkage Table (PLT). The PLT obtains addresses of external definitions from the Global Offset Table (GOT). The GOT entry for a symbol contains the absolute virtual address of a corresponding definition.

Our linker (discussed next) nullifies the GOT entries for unavailable external functions. The code in Figure 7 exploits this phenomenon to implement FAC. It checks if the GOT entry for a target function (foo) contains NULL. Success implies non-availability of definition and execution skips the FAC compound. Conversely, failure implies availability of a valid definition and other instructions within the FAC compound are executed. Note, once again, that there is a semantic difference between a NULL assigned function pointer and a NULL containing GOT entry. The former is an existing and valid programming concept, while the later is not under control of contemporary programming languages. Thus, nullification of GOT entries can safely imply non-availability.

```
        call    .L2
.L2:
        popl    %eax
        addl    $_global_offset_table_+[.-.L2], %eax
        cmpl    $0, foo@GOT(%eax)
        je      .L1
        call    foo
        … # other code inside compound FAC
.L1:
        … # code outside FAC compound.
```

Figure 7: FAC assembly for the source in Figure 3. Lines 1 – 4 set up access to the GOT. Lines 4 – onwards implement FAC.

An issue arises due to referential completeness checks enforced by the static link editor. These checks can hinder successful compilation in the absence of a utility library. Nonetheless, we may reasonably expect all pieces of a program to be available during initial development for purposes such as verification. Otherwise, dummy implementations can be provided for the target functions in a dummy library to overcome this problem. For proof of concept, we currently patch post-compilation assembly to instrument FAC code wherever required. We are in the process of modifying the GCC backend to automate this procedure.

## 6.2 Program Loader/Linker

Since contemporary loaders rigorously require availability of all utility libraries at startup, they never allow programs to benefit from runtime adjustments through FAC. To address this problem, we modified the program loader to ignore missing libraries assuming that resultant undefined references are covered through FAC constructs. If undefined references are left outside FAC compounds, the program will potentially crash. However, this is like attempting to access a `dlsymed` symbol without verifying its validity [10]. It is therefore reasonable that a programmer takes care in this regard.

Our modified linker tolerates undefined external references and nullifies GOT entries of unavailable external functions during load-time resolution. All other types of unavailable external symbols, such as global variables, are ignored assuming FAC work-around. Since we require that FAC constructs circumvent all unavailable external elements, the runtime dynamic linker should never have to be invoked on undefined references. Consequently, failure of lazy symbol resolution implies an error condition and regular action is taken. A prototype for our modified linker/loader [6] is available for download at http://blandings.cs.vt.edu/~joy.

## 6.3 Weak Aliasing

FAC may be implemented using weak aliasing [10] techniques along with position independent compilation (PIC). Since weak references need not be compulsorily resolved, C supports FAC-like checks on them. The generated code is similar to FAC [Figure 7]. Weak aliasing and PIC can, therefore, be used to obviate compiler modifications discussed in section 6.1. However, there are two issues with this approach. Firstly, weak references entail extra preprocessor constructs (such as `#pragma weak`). Secondly, using PIC can result in extra assembly than necessary since it affects an entire source file rather than a single expression. Nevertheless, our method can be seen as an extended and optimized version of this approach.

## 7 Generic Availability Checks

From the implementation viewpoint, FAC is the simplest and most efficient of possible availability checks. For instance, performing availability checks on variables would require a new programming construct since 'if' checks on them have traditional semantics. Besides, functions are the atoms of procedural codes. Hence, it is unlikely that an external library or module would not contain functions. Accesses to variables can, hence, be covered by FACs on co-defined functions. Performing availability checks on library names would require transferring control to the system loader. The loader can verify availability through lookups on the list of libraries in the executable's scope. However, such a lookup would incur needless overheads compared to FACs.

# 8 Preliminary Evaluation

## 8.1 Programming Aspects

A comparison of the codes in Figure 2 and Figure 4 provides an idea of FAC benefits from the programming perspective. The former comprises nearly twice as many lines of code as the later. In our opinion, the FAC code is also more intuitive and readable due to simple correspondence with the established 'if' construct. For advanced measurements, we modified DILLO to incorporate automatic adjustments to library support for displaying GIF, JPEG, and PNG images (a total of 3 external shared objects). While a FAC based implementation *lessened* the original code by 4 lines, a dynamic library (`libdl`) based one resulted in an extra 23 lines. Furthermore, from an overall code-base viewpoint, FAC obviated the necessity of `libdl` (10 Kbytes) and the associated header `dlfcn.h` (240 lines). To remain unbiased towards either approach, we minimized source-level changes to the original DILLO code. Unmodified original sources for JPEG, GIF, and PNG support were compiled into corresponding shared objects. Source alterations were restricted to incorporating dynamic adjustments in core executable code only. *Working versions of our FAC-based and libdl-based DILLO codes can be accessed at the pre-mentioned website.*

Due to limitations of our current prototype implementation (especially compiler modifications), a FAC-based XMMS is as yet incomplete. Additionally, XMMS sources (including plugins) adhere to a dynamic library oriented design. For fair comparison, we need to unravel all such adherence and reprogram everything along a FAC-based path. However, due to the pervasiveness of complex plugin-based code throughout XMMS, this task is taking more time than expected.

## 8.2 Flexibility Aspects

Along with our modified linker/loader, FAC allows a program to flexibly adjust to availability of utility libraries and associated program entities. Peripheral features, implemented as utility libraries, can be simply added to or removed from a specified directory to add or prune respective functionality without rebuilding. Our website contains a demonstration on DILLO that flexibly adjusts to arbitrary addition or removal of utility libraries for JPEG, PNG, and GIF support.

A FAC-based approach requires restart to incorporate a previously missing library. This is because utility libraries are loaded once only during startup. However, many applications using libdl (XMMS, for instance) also load and link dynamic libraries once during initialization only. This limitation is, therefore, not too disturbing. Again, a FAC-based approach does not offer programmer control over explicit linking of symbols. This factor is a bottleneck for handling of scenarios involving alternate libraries for a single purpose with each exposing similar interfaces. For instance, XMMS enumerates several alternate plugins for identical purposes and allows the user to select among them at runtime. All the plugins, which use similar symbols, must therefore be loaded with user control over links to the main executable. This feature has been one of the prime reasons for complications in our attempts at incorporating FAC support into XMMS.

Nevertheless, in defense, we note that our primary vision for FAC (see section 1) is automatic adjustment to addition or pruning of *different features* as opposed to enumeration of multiple availabilities. We reason that libraries for different purposes would, largely, incorporate dissimilar usage patterns, signatures, and symbols.

## 8.3 Performance Aspects

Runtime overheads due to FAC are nearly the same as in the regular use of dynamic libraries (`libdl`). Moreover, as mentioned earlier, available optional features and services are often tallied during initialization only thus abating recurring runtime costs. Hence, we restrict performance evaluations to initialization costs only.

Figure 8 depicts initialization costs for FAC and dynamic library implementations of DILLO against the original version. The experiments were run on GNU/Linux over a 2GHz Athlon64 processor (1GB RAM) in x86 mode. Unmodified sources for JPEG, PNG, and GIF support were compiled into shared objects. APIs in `libdl` were used for the dynamic library implementation. To restrict analyses to FAC only, the system loader was used to load all optional libraries. Thus, a FAC version had to be realized using weak aliasing and PIC (section 6.3). The original version was reconfigured and recompiled with different degrees of support. A wrapper program was used to start the main application using the system command. Times were measured (using gettimeofday) just before the system command in the wrapper and then inside the browser main code after all available peripherals had been loaded. Lazy linking was used in all cases to minimize initialization costs.

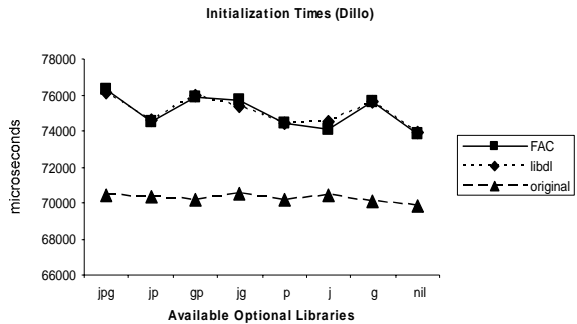

Figure 8: A FAC-based approach did not result in discernible performance changes with respect to a dynamic library (libdl) based one. In the figure, 'j' stands for jpeg, 'p' for png, and 'g' for gif. 'jpg' implies available support for all optional libraries. 'jp' implies availability of jpeg and png utilities and so on. Optional library combinations are listed in the order of decreasing total sizes from left to right.

The original version showed a slight linear decrease in costs due to decreasing size. We attribute the variations to lazy linking against other utility libraries such as libc. Both the FAC and libdl based cases incurred a greater cost due to additional dynamic loading and linking. However, there was no discernible difference in the extra costs incurred in either approach. Thus, FAC did not add any performance overheads with respect to traditional dynamic libraries. Moreover, both FAC and libdl based cases showed greater and, seemingly, random variation in terms of costs vs. size. We attribute this randomness to unpredictability of lazy linking as documented in [11]. Interestingly, the maximum variations were cause due to the GIF support library. We are currently investigating the reasons for this. More details and inferences can be obtained at our website.

## 9  Summary and Ongoing Work

We have shown that function availability checks can be used in programs to achieve automatic adjustments to presence or absence of segments of code. FAC facilitates customization of application functionality through addition or removal of optional utility libraries, while obviating complete re-building. In doing so, FAC lessens development complexity, crudely measured as lesser lines of code, with respect to prevalent mechanisms in the domain of compiled procedural languages such as C. Ongoing work involves, mainly, completing compiler modifications for FAC and implementing FAC-based XMMS plugins. Porting FAC to more architectures, languages (C++), and operating systems is another focus.